\documentclass[preprint]{article}

\usepackage{algpseudocode}

\usepackage[nonatbib]{neurips_2023}

\usepackage[utf8]{inputenc} 
\usepackage[T1]{fontenc}    
\usepackage{hyperref}       
\usepackage{url}            
\usepackage{booktabs}       
\usepackage{amsfonts}       
\usepackage{nicefrac}       
\usepackage{microtype}      
\usepackage{xcolor}         
\usepackage{natbib}
\usepackage{listings}
\usepackage{algpseudocode}
\usepackage{graphicx}
\usepackage{mathtools}
\usepackage{array}

\title{Optimizing sDTW for AMD GPUs}

\author{%
  Daniel Latta-Lin \\
  Department of Computer Science\\
  University of Washington\\
  Seattle, WA 98195 \\
  \texttt{danielin@uw.edu} \\
  \And
  Sofía Isadora Padilla Muñoz \\
  Department of Computer Science\\
  University of Washington\\
  Seattle, WA 98195 \\
  \texttt{sofisa16@uw.edu} \\
}

\begin{document}
\graphicspath{./images/}

\maketitle

\begin{abstract}
Subsequence Dynamic Time Warping (sDTW) is the metric of choice when performing many sequence matching and alignment tasks. While sDTW is flexible and accurate, it is neither simple nor fast to compute; significant research effort has been spent devising parallel implementations on the GPU that leverage efficient memory access and computation patterns, as well as features offered by specific vendors and architectures (notably NVIDIA's).

We present an implementation of sDTW on AMD hardware using HIP and ROCm. Our implementation employs well-known parallel patterns, as well as lower-level features offered by ROCm. We use shuffling for intra-wavefront communication and shared memory to transfer data between consecutive wavefronts. By constraining the input data to batches of 512 queries of length 2,000, we optimized for peak performance the width of reference elements operated on by a single thread.
\end{abstract}

\maketitle

\section{Introduction}
Real-time signal data has become more and more abundant with the rise of high frequency trading, nanopore gene sequencing and audio signal processing. Quickly recognizing patterns in this noise is incredibly important to react to market trends and also in order to keep up with advances in hardware improvements for gathering data. Along with the increase in real-time data acquisition, GPU architectures are emerging that allow for a new programming paradigm that promises an order of magnitude or more of increased efficiency. To tackle this problem and take advantage of new GPU architectures we introduce a new sub-sequence Dynamic Time Warping algorithm built to leverage parallelism on AMD GPUs. 

\section{Background}
Dynamic Time Warping (DTW) has long been a tool that allows for similarity comparisons between streams of signals or time series data. Originally used for speech recognition\cite{speech-recog}, it allows for the comparison of similarity between signals that might be stretched at different rates across temporal space or shifted. What matters is not whether there's a perfect alignment or match between the signals, but whether they are similar (and where, that is, at which points), as quantified by some metric that measures the cost of converting (or \textit{aligning}) one sequence to the other via insertions, deletions, and substitutions (\cite{sdtw}).

The alignment computed with DTW can be regarded to be \textit{global}, where the two series are aligned across their full length, or a \textit{subsequence} alignment, where one of the series (the \textit{query}) is typically shorter than the other one (the \textit{reference}) and the objective is to find the subsequence (a segment) of the reference that the query aligns with or fits the best. In this paper, we are concerned with a variant of the DTW problem, \textit{subsequence DTW} (sDTW), which computes a subsequence-type of alignment.

Before DTW, most similarity comparisons used to employ the Euclidean distance metric. Computation of this metric is known to be efficient and simple to implement. However, this metric suffers from a number of problems that reduce its accuracy significantly in classification tasks (feature or pattern identification) and clustering tasks (grouping based on features) (\cite{euclidean}).

While DTW is the method of choice for similarity comparisons, precisely because it addresses the shortcomings of the Euclidean metric, it is far more computationally expensive and more involved to implement. With the advent of massively parallel processors, many researchers set out to design and implement efficient DTW-based algorithms that leverage the growing computing power of GPUs. We discuss some of the most notable works in section \ref{sec:prior-works}. To get a sense of its complexity, and to appreciate the many ways in which it lends itself to parallelization, let us look closer at the formulation of the problem and the DTW and sDTW algorithms.

The problem asks, given 2 signals \textit{X} and \textit{Y} of lengths \textit{N} and \textit{M}, respectively, and a value cost function \textit{C} what is the mapping, or "path", between \textit{X} and \textit{Y} that results in the minimum overall cost? If we arrange the elements of \textit{X} and \textit{Y} such that there's 1 row per element of \textit{X} and 1 column per element of \textit{Y}, we obtain a \textit{NxM} matrix. On this matrix, DTW establishes a correspondence between each element of \textit{X} and a set of adjacent elements of \textit{Y}. These correspondences form a path (a \textit{warp path}) from one corner of the matrix to the opposite corner, advancing one element at a time through \textit{X} or \textit{Y} or both simultaneously. DTW can then be understood as the search for the optimal warp path.

More specifically, the search for the optimal warp path follows a dynamic programming strategy that maintains the aforementioned matrix, filling it in from one corner to the opposite corner, with the value of a cell \textit{(i,j)} being the sum of the minimum accumulated distance between $X_i$ and $Y_j$ so far and the minimum value of the 3 neighboring cells, all of which have been computed already and represent the cost of 3 alternative paths that align $X_i$ and $Y_j$ by insertion of elements ($X_i$ is aligned with a gap in $Y$), deletion, or inaction ($X_i$ and $Y_i$ match). Formally, the cost function $D(i,j)$ is given by the following recurrence relation:
\begin{equation}
    D(i,j) \coloneq min\{D(i-1,j), D(i,j-1), D(i-1, j-1)\} + d(X_i, Y_j)
\end{equation}
where $d$ is a measure of distance between the 2 given elements. After the matrix is populated completely, the optimal warp path is found by walking back from the minimum valued tile in the last row back to the opposite corner of the matrix, following the neighboring cell \textit{(i,j), (i-1,j), (i,j-1)} with the smallest accumulated cost. Depending on the type of alignment sought after, the matrix is initialized differently and the walk-back pass proceeds from a different starting point and ends where the query ends.

Dynamic time warping has found many use cases outside of audio processing and speech recognition. This includes the identification of economic trends in financial time series; the detection of certain seismic patterns during monitoring of seismic activity (\cite{seismic}); the recognition of specific individual gestures in motion-captured data as used in computer animation (\cite{mocap}); the identification of subsequences in protein and DNA sequences; and the detection of anomalies and unusual activity in network traffic (\cite{network}).

\section{Prior Works}
\label{sec:prior-works}

DTW has proven to be one of the most accurate and flexible metrics for measuring the similarity of two data series and computing optimal alignments. However, unlike other metrics, like Euclidean distance, computing DTW is far from being fast. With the advent of general purpose computing on GPUs, however, great strides have been made toward efficient and fast sDTW solutions that leverage efficient memory access patterns and make use of warp intrinsics that accelerate the computation.

One of the earliest parallelizations on the GPU of a DTW algorithm is due to Hundt et al. (\cite{cudadtw}). Their paper describes parallelization strategies at the thread-, block-, and batch-levels for a constrained variant of DTW. Of interest to our purposes is their use of each thread block for computing a single cell of the matrix and their extensive use of share memory. Future work by other researchers found weaknesses in these two aspects of their implementation and proposed more efficient alternatives. 

Schmidt et al. (\cite{cudaDTW++}) achieved a speedup of a couple of orders of magnitude in subsequence search with \textit{cuDTW++} (in the context of ECG analysis), compared to the best-performing CPU and GPU solutions at the time (UCR-Suite and cudaDTW), by using low-latency, intra-warp communication intrinsics such as \textit{\_\_shfl\_down\_sync\(\)} and \textit{\_\_shfl\_up\_sync\(\)} instead of global and shared memory, a novel partitioning scheme of the DTW matrix, with a corresponding work distribution among threads; and an additional early termination condition that avoids unnecessary, useless work.

Sadasivan et al. (\cite{DTWAX}) presented \textit{DTWax}, a high-throughput solution that reorders the reference offline for promoting coalesced global memory access, pre-processing the input batch by normalizing it, and making extensive use of Fused-Multiply-Add instructions of the Matrix-Multiply-Accumulate pipe to perform addition in the computation of the cost function. A number of other smaller optimizations are employed as well; of notable interest to us was their use of loop unrolling for reducing branch divergence.

\section{Methodology}
\label{method}
The goal of our work is to generalize these algorithms to the ROCM framework. ROCm supports float16 values along with APIs that allow for the manipulation of two float16 values at once. In addition to leveraging lightweight data-types we want to maximize the parallelism between executing threads. This means minimizing the shared memory dependencies between threads and allowing for efficient information sharing with as little blocking as possible. To support this concurrency we leverage warp shuffling to transfer information between columns of our DTW matrix.

To arrive at our final implementation, our solution evolved through a number of iterations. The normalizer, for example, started out as a collection of kernels that performed, in stages, the different operations involved in the standardization of a single query; on the other hand, the final version of the normalizer is an all-in-one kernel that processes an entire batch at once, that uses the well-known pattern of parallel reduction and the principle of thread coarsening, and that tries to promote coalesced memory accesses. Similarly, the sDTW kernel started out performing the alignment of a single query against the reference; multiple calls to the kernel used to be required to process an entire batch and large pieces of shared memory had to be maintained in sync across all working threads; the final version, however, is a high-throughput kernel that runs large batches in parallel on separate wavefronts and isolates threads within wavefronts.

The development of our solution was accompanied by the development of a test dataset generator written in Python. This data generator uses the \textit{make\_cylinder\_bell\_funnel} time-series generator of the \textit{pyts.datasets} package to produce references and queries of specified lengths that served as unnormalized inputs to our C++ CUDA modules. Along with these inputs, our test dataset generator implements versions of the normalizer and sDTW algorithms for execution on the CPU, with the strict purpose of producing the expected output of a CUDA sDTW batch run for correctness evaluation; naturally, producing these expected outputs on the CPU is a time-consuming process.

\section{Implementation}

Our implementation is comprised of two C++ CUDA modules. The first one is the \textit{normalizer}; this module exposes a \textit{runNormalizer} function that is given a batch size (the number of queries included in the batch), a query size, a pointer to a CPU-side buffer populated with the whole batch of queries, and a pointer to a GPU-side buffer, where the normalizer will place the normalized batch. The second module orchestrates the calls to the normalizer (to normalize both the reference and the batch of queries), makes the required CPU- and GPU-side allocations, and implements \textit{sDTW} in the \textit{runSDTW} function. 

This section is organized as follows. We start by describing our development environment. We then describe in detail the normalizer program and its most important component, the normalization HIP kernel. We proceed to explain the implementation details of our sDTW HIP kernel, arguably the most important part of our work. We end this section by explaining how we tested the implementation.

\subsection{Normalizer implementation}
\label{normalizerimpl}

As explained before, both the query and reference are series of floating-point values. In practical applications, the reference series or the query may be the result of measurements or data captures that vary in scale; these differences, for example, may account for the use of different units of measurement or different volumes of recorded data. Differences in scale introduce a number of problems at the time of analysis; for example, high-level trends or patterns that both the reference and the query possess may go unnoticed if the magnitude of the series' elements is large in one and small in the other. Another common problem is that of floating-point numerical instability caused by the use of extremely large and extremely small floating-point values.

To avoid these problems, both the reference and the query are typically pre-processed to be on a common scale, by normalizing their values. In particular, our implementation standardizes the input so that the dataset has a mean of 0 and a standard deviation of 1.

\begin{equation}
\label{standardization}
z = \frac{x-\bar{x}}{S}
\end{equation}

where $\bar{x}$ denotes the sample mean of the original series and $S$ is its standard deviation.

With the purpose of performing multiple alignment queries in a single call to the sDTW kernel, our normalization kernel processes, at once, an entire batch of query series. All the queries are expected to be stored contiguously in global memory, one after the other, with no gaps, delimiters or extra metadata.

By design, one block is assigned to each query in the batch. Ideally, each thread in the block would normalize one element of the query (making the algorithm embarrasingly parallel), with all of them collaborating to compute the sample mean and standard deviation, but each block had a limit of 1024 threads in our development environment and each query was assumed to be 2000 elements long. Therefore, we employed the thread coarsening principle to give each thread up to 2 elements to normalize.

Mathematically and conceptually, the calculation of the sample mean and standard deviation is performed as in the CPU-side code of cuDTW++:

\begin{lstlisting}
sum /= n;
sumSq = sumSq/n - sum*sum;
\end{lstlisting}

where $sum$ denotes the sum of all the query elements, $sumSq$ denotes the sum of their squares, and $n$ is the number of elements. Threads in the block collaborate to compute the sum and sum of squares. Using an array in shared memory (of dynamic size specified at kernel launch time), threads store the partial sum and partial sum of squares of their assigned elements. For promoting coalesced access to these sums later on, shared memory is partitioned into 2 segments: the first half stores the partial sums contiguously, whereas the second half stores the partial sums of squares.

Once all the partial sums are computed and stored in shared memory, parallel reduction is done to iteratively compute further partial sums until, ultimately, a single sum and sum of squares are obtained (it is worth noting that there's nothing novel in the way we do parallel reduction here).

With the sum of elements and their squares computed, it is up to the first thread in the block to calculate the sample mean and standard deviation. At this point, since the partial sums in the shared memory array are no longer of use, we reuse the first two elements to store the sample mean and standard deviation; this way, these two values become visible to the rest of the threads in the block.

The last step in the normalization process involves all threads waiting for the first thread to compute and write the sample mean and standard deviation in shared memory, and then each of them applying the standardization operation of equation \ref{standardization} to their assigned elements.

\subsection{sDTW implementation}
Our implementation is inspired by and leverages many of the same design goals as \cite{DTWAX}. The goals of our sDTW implementation is to maximize parallelism while also making operations as lightweight as possible so we can achieve a high throughput. To achieve lightweight operations we first preprocess the Q query float32 values, which represent Z equal length queries concatenated, along with the N reference float32 values to which the queries will be compared. The transformation is from float32 to float16 data types. ROCm allows these values to be packed together in pairs in a datatype called \textit{\_\_half2} where each float16 value is considered a \textit{\_\_half}. The output of the initial preprocess step are 2 new buffers. One contains (Q + 1)/2 \_\_half2 values while the other contains (N+1)/2 \_\_half2 values.

We calculate the path through the cost matrix in parallel. Each thread is responsible for a section of the reference string which we will refer to as a \textit{segment}. Threads proceed to calculate values for their segments row by row but each segment will have dependencies on the previous segments values. To propagate previous values, one approach might be to calculate diagnol values and place these values in shared memory where the next thread can easily find previous values for cost calculation. This implementation proved to be complex and also introduced many more \textit{\_\_sync\_threads} calls which reduce the concurrency of our code. Instead of using shared memory to pass values between threads we use ROCm's \textit{\_\_shufl\_up} command which performs a parallel shuffling of values across a warp. Given 2 threads 1,2 which are processing adjacent segments, thread 1 will propagate the last value in it's segment so that it may be used as the left side of thread 2's cost calculation in the next iteration. Each thread keeps 2 local buffers of segment width. The first buffer represents values from last iteration while the second is the buffer that is currently being filled. At the end of each iteration the buffer pointers are flipped. This allows each thread to work in almost pure isolation relying only on efficient warp shuffles to propagate information for cost calculation.

\begin{figure}
        \includegraphics[width=\linewidth, height=10cm]{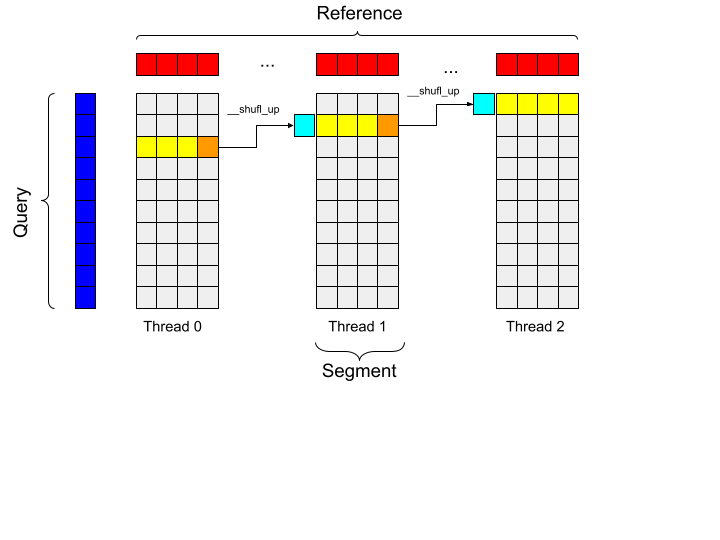}
        \label{fig:gradient-ff60}
        \caption{On each iteration threads calculate the cost values for the next row in their segment. The right most value is then propagated up to be used by the next thread in the following iteration.}
\end{figure}

On each iteration, active threads process a new row of query data. Inactive threads are threads that do not yet have left-side information propagated to calculate their segment cost. Each iteration can be seen as calculating a new diagnol on the cost matrix. Once the diagnol that is being calculated has reached the end of the query string (the bottom of the cost matrix) we may begin extracting the minimum cost. We avoid waiting to calculate the minimum cost of the bottom row of the cost matrix by propagating values from the bottom row of the cost matrix as they are being generated. Once a thread has reached the bottom of the cost matrix it calculates the smallest pair of floating point values in the segment using the pairwise minimum finding operation \textit{\_\_hmin2}. Once the segment local minimum \_\_half2 value is calculated it is propagated to the next thread using another \_\_shufl\_up command. This allows the minimum aggregation to be carried out across iterations at the same time the final values are calculated.

Once the final thread has reached the bottom row the minimum cost value for that wave front has been calculated. If the reference string is too large to be processed by a single wavefront we must save this wavefront minimum in shared memory to be propagated to the next wavefront for further aggregation. Since shuffling is limited by wavefront width we must also propagate the right side column of the cost matrix from the previous wavefront when starting a new wave front. This is done by updating a shared memory buffer which represents the last segment values for the final thread in a wavefront. This shared memory is only touched by this final thread for updates and is read by the initial thread in the wavefront. To avoid conflicts we again maintain two buffers one for reading and the other for writing. These buffers are switched at the end of the wavefront allowing for a single \_\_sync\_threads call per iteration.

To support large batches of queries we simply allocate a compute block for each query. These blocks will be the size of a wavefront and each block will be able to calculate it's own section from the query buffer in parallel. Results are then written back out to global memory in parallel.

\begin{figure}
        \includegraphics[width=\linewidth, height=10cm]{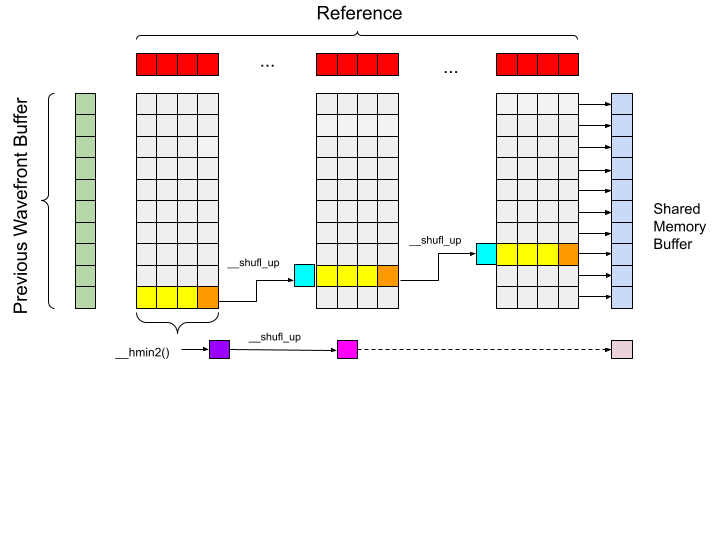}
        \label{fig:gradient-ff60}
        \caption{Once a thread is finished with it's segment the found minimum result for the segment is propagated up for further enrichment. The shared memory buffer allows for communication between a completed and newly started wavefront.}
\end{figure}

\section{Results}

We evaluated the results of our implementation along two dimensions: correctness and performance. As described in section \ref{method}, we used our test dataset generator to produce the expected output of a GPU-accelerated sDTW run based on the output produced by a CPU-based sequential version of the algorithm; this CPU-side version is much more straightforward, so we are confident in the correctness of its output. Leveraging the \textit{wb} library, we compared the output of our CUDA-based sDTW solution against the expected output.

In terms of performance, we evaluated our solution using two metrics: execution time and throughput in gigasamples per second (Gsps). The latter was computed using the following formula:

\begin{equation}
  gigasamplesPerSecond \coloneq floatsProcessed / (milliseconds * 1e9 / 1000)
\end{equation}

where $floatsProcessed$ is the total number of floating-point values in all queries in the batch and $milliseconds$ measures the execution time of the kernel in question in milliseconds.

Given a batch of 512 queries of 2,000 unnormalized floating-point elements each and a reference series of size 100,000, we computed the average execution time and average throughput of the two kernels based on 10 runs (preceded by 2 cold runs meant for warming up the GPU). The results are displayed in table \ref{tab:stats},

\begin{table}[]
    \centering
    \begin{tabular}{|l|l|l|}
        \hline
        & \textbf{Throughput (Gsps)} & \textbf{Execution time (ms)} \\ \hline
        sDTW kernel & 0.000926544 & 11036.5 \\ \hline
        Normalizer kernel & 4.81973 & 0.0214238 \\ \hline
    \end{tabular}
    \caption{Average performance statistics based on 10 runs}
    \label{tab:stats}
\end{table}

We also experimented with different segment widths. This has the affect of coarsening each thread. We found that for our batch size of 512 queries of 2000 values and reference string of length 100,000 performance peaked at around a segment width of 14 with performance degradation after. Testing was done using an average of 10 runs after an initial 2 warm up runs of the kernel.

\begin{figure}
        \includegraphics[width=\linewidth, height=7cm]{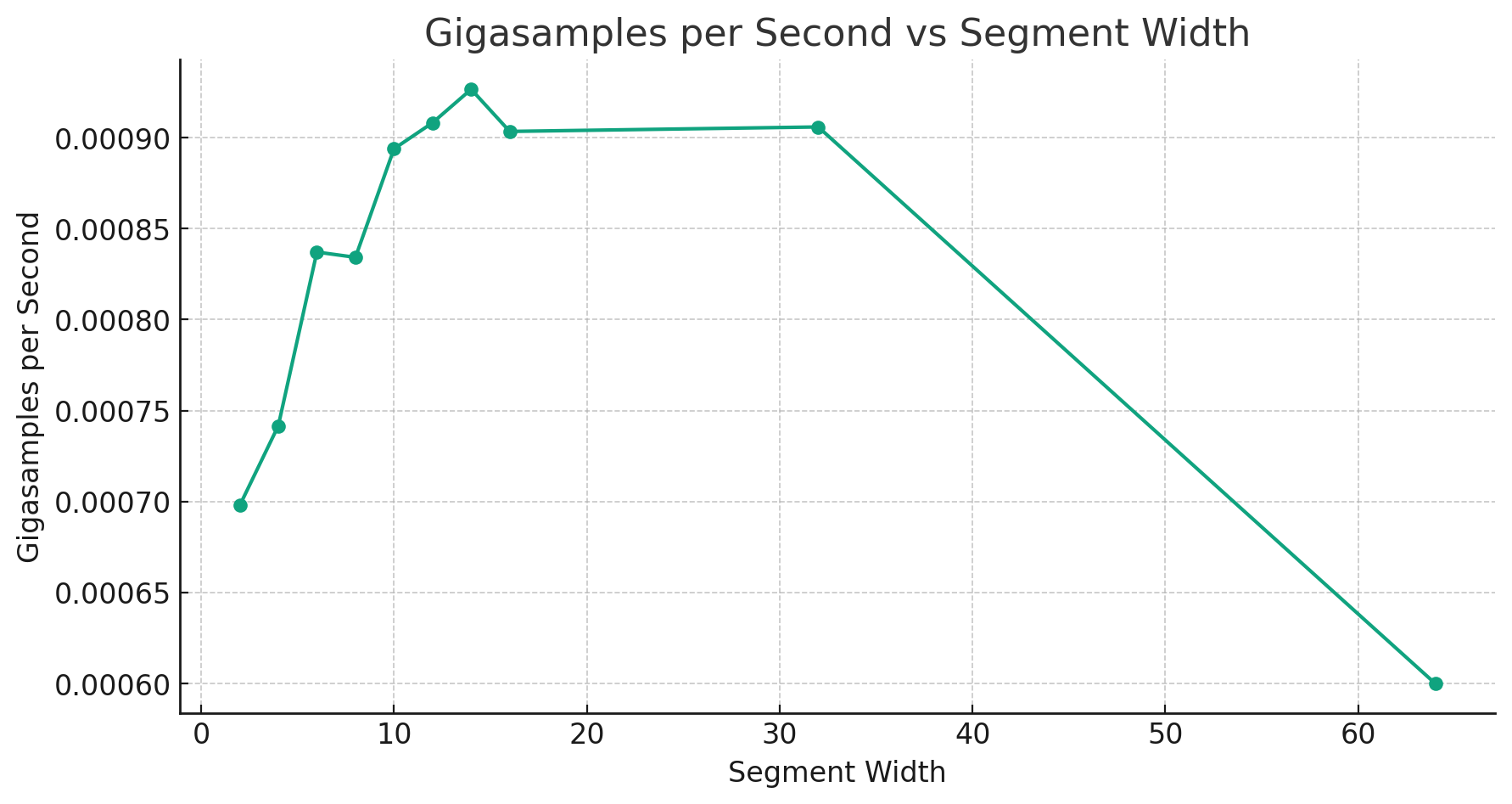}
        \label{fig:gigsample}
        \caption{Initially increasing segment width improves performance but any gains are soon lost once segment width grows past 14 values wide.}
\end{figure}

Coarsening threads to process 14 reference values at a time helped improve performance by ~30\% from the minimum segment width of 2 values.

\section{Conclusion}

In this paper, we described an alternative parallel implementation of the subsequence Dynamic Time Warping (sDTW) algorithm on AMD GPUs. For a given input batch of 512x2,000 queries and a reference of size 100,000, our sDTW kernel achieves a throughput of 9.26e-4 gigasamples per second on average; our normalizer kernel, on the other hand, achieves a throughput of 4.81 gigasamples per second.

Our sDTW kernel operates on float16 values using the pairwise GPU intrinsics supported by ROCm. Our sDTW computation is done isolated per thread using shuffling for intra-warp communication and shared buffer memory to transfer state between consecutive wavefronts. We also optimize our segment width to coarsen each thread for peak performance given the input constraints. The normalizer performs z-normalization to eliminate many of the problems associated with differences of scale in the input data; it processes one query per block, where each thread is coarsened to process multiple elements of the query, using shared memory to collaborate on computing the necessary statistics using the well-known parallel reduction pattern.

\section{Discussion}
It seems likely that there are even more gains to be drawn from further quantization. Other work on LLMs have focused on extreme clipping and quatnization in order to get large performance gains with little degradation in output quality \cite{bitLLM}. An approach we would like to try but did not have the time is to further quantize the fp16 values down to uint8. This approach would first involve generating a codebook based on the reference string. To produce the codebook we would like to get the distribution of floating point values and then evenly divide the bulk of the distribution across uint8 values clamping any outliers to the extreme values. Another idea we did not get time to explore was leveraging early pruning of values that have a large separation in distance. When performing cost calculation we can perform the initial subtraction and then if the values seem to qualify as "far" apart we may assume that the tile does not contribute to the path and simply return an infinite value (INF) instead of performing multiplication. These INF tiles would further reduce the number of multiples performed downstream.

Overall even with the given work it seems there is more opportunity for speed up using pruning techniques and even lighter weight datatypes but with our implementation we have shown that previous work in this field is easily portable to new and emerging architectures.
\bibliography{citations}

\begin{thebibliography}{10}
\providecommand{\natexlab}[1]{#1}
\providecommand{\url}[1]{\texttt{#1}}
\expandafter\ifx\csname urlstyle\endcsname\relax
  \providecommand{\doi}[1]{doi: #1}\else
  \providecommand{\doi}{doi: \begingroup \urlstyle{rm}\Url}\fi

\bibitem[Myers(1980)]{speech-recog}
C.~Myers.
\newblock A comparative study of severa l dynamic time-warping algorithms for speech recognition.
\newblock 1980.

\bibitem[Candan et~al.(2012)Candan, Rossini, Wang, and Sapino]{sdtw}
K.~Sel\c{c}uk Candan, Rosaria Rossini, Xiaolan Wang, and Maria~Luisa Sapino.
\newblock sdtw: computing dtw distances using locally relevant constraints based on salient feature alignments.
\newblock \emph{Proc. VLDB Endow.}, 5\penalty0 (11):\penalty0 1519–1530, jul 2012.
\newblock ISSN 2150-8097.
\newblock \doi{10.14778/2350229.2350266}.
\newblock URL \url{https://doi.org/10.14778/2350229.2350266}.

\bibitem[Ratanamahatana and Keogh(2004)]{euclidean}
Chotirat~Ann Ratanamahatana and Eamonn Keogh.
\newblock Making time-series classification more accurate using learned constraints.
\newblock pages 11--22, 2004.
\newblock \doi{10.1137/1.9781611972740.2}.
\newblock URL \url{https://epubs.siam.org/doi/abs/10.1137/1.9781611972740.2}.

\bibitem[Kumar et~al.(2022)Kumar, Legendre, Zhao, and Chao]{seismic}
Utpal Kumar, Cédric.~P. Legendre, Li~Zhao, and Ben~F. Chao.
\newblock {Dynamic Time Warping as an Alternative to Windowed Cross Correlation in Seismological Applications}.
\newblock \emph{Seismological Research Letters}, 93\penalty0 (3):\penalty0 1909--1921, 03 2022.
\newblock ISSN 0895-0695.
\newblock \doi{10.1785/0220210288}.
\newblock URL \url{https://doi.org/10.1785/0220210288}.

\bibitem[Switonski et~al.(2019)Switonski, Josinski, and Wojciechowski]{mocap}
A.~Switonski, H.~Josinski, and K.~Wojciechowski.
\newblock {Dynamic time warping in classification and selection of motion capture data}.
\newblock \emph{Multidim Syst Sign Process}, 30:\penalty0 1437–1468, 2019.

\bibitem[Zhan et~al.(2020)Zhan, Xu, Luo, and Li]{network}
Peng Zhan, Haoran Xu, Wei Luo, and Xueqing Li.
\newblock A novel network traffic anomaly detection approach using the optimal $\varphi$-dtw.
\newblock In \emph{2020 IEEE 11th International Conference on Software Engineering and Service Science (ICSESS)}, pages 1--4, 2020.
\newblock \doi{10.1109/ICSESS49938.2020.9237659}.

\bibitem[Hundt et~al.(2014)Hundt, Schmidt, and Schömer]{cudadtw}
Christian Hundt, Bertil Schmidt, and Elmar Schömer.
\newblock Cuda-accelerated alignment of subsequences in streamed time series data.
\newblock In \emph{2014 43rd International Conference on Parallel Processing}, pages 10--19, 2014.
\newblock \doi{10.1109/ICPP.2014.10}.

\bibitem[Schmidt and Hundt(2020)]{cudaDTW++}
Bertil Schmidt and Christian Hundt.
\newblock cudtw++: Ultra-fast dynamic time warping on cuda-enabled gpus.
\newblock \emph{European Conference on Parallel Processing}, 2020.
\newblock URL \url{https://link.springer.com/chapter/10.1007/978-3-030-57675-2_37}.

\bibitem[Sadasivan and Stiffler(2023)]{DTWAX}
Harisankar Sadasivan and Daniel et~al. Stiffler.
\newblock Accelerated dynamic time warping on gpu for selective nanopore sequencing.
\newblock \emph{bioRxiv}, 2023.
\newblock URL \url{https://doi.org/10.1101/2023.03.05.531225}.

\bibitem[Ma et~al.(2024)Ma, Wang, and Ma]{bitLLM}
Shuming~Ma Ma, Hongyu Wang, and Lingxiao et~al. Ma.
\newblock The era of 1-bit llms: All large language models are in 1.58 bits.
\newblock \emph{PrePrint}, 2024.

\end{thebibliography}
\end{document}